\documentstyle[12pt,openbib]{article}
\large
\begin{document}

\title
{q-deformed pairing-vibrations}

\author{S. Shelly Sharma\\
Depto. de F\'isica,
Universidade Estadual de Londrina,\\
86051-970 - Londrina, Parana, Brazil\\
N. K. Sharma\\
Depto. de Matem\'atica, Universidade Estadual de Londrina,\\
86051-970 - Londrina, Parana, Brazil} \date{} \vspace{2 true cm}
\maketitle

\begin{abstract}
Boson creation operators constructed from linear combinations of q- deformed
zero coupled nucleon pair operators acting on the nucleus (A,0), are used to
derive pp-RPA equations. The solutions of these equations are the pairing
vibrations in (A${\underline{+}}$2) nuclei. For the $0^+_1$ and $0^+_2$
states of the nucleus $^{208}$Pb, the variations of relative energies and
transfer cross-sections for populating these states via (t,p) reaction, with
deformation parameter $\tau$ have been analysed. For $\tau=0.405$ the
experimental excitation energy of 4.87MeV and the ratio $\frac{\sigma(0^+_2)
}{\sigma(0^+_1)}=0.45$ are well reproduced. The critical value of pairing
interaction strength for which phase transition takes place, is seen to be
lower for deformed zero-coupled nucleon pair condensate with $\tau$ real,
supporting our earlier conclusion that the real deformation simulates the
two-body residual interaction. For $\tau$ purely imaginary a stronger
pairing interaction is required to bring about the phase transition. The
effect of imaginary deformation is akin to that of an antipairing type
repulsive interaction.

Using deformed zero coupled quasi-particle pairs, a deformed version of
Quasi-boson approximation for $0^+$ states in superconducting nuclei is
developed. For the test model of 20 particles in two shells, the results of
q-deformed boson and quasi-boson approximations have been compared with exact
results. It is found that the deformation effectively takes into account the
anharmonicities and may be taken as a quantitative measure of the
correlations not being accounted for in a certain approximate treatment.

\vspace{0.8 true cm} PACS Number: 21.60.Fw, 21.60.Jz, 21.10.Re
\end{abstract}

\Roman{section}

\section{Introduction}

\hspace{.25in}

The notion of quantum groups has attracted a lot of attention over the last
few years. The quantization by deformation was studied earlier by Bayen
et. al \cite{Bay78}. Drinfeld \cite{Dri86} generalized this idea to quantize
classical Lie algebras so as to construct non-commutative Hopf algebra
structure. In particular the quantum group $SU_q$(2), the q-analog of $SU$%
(2), has been extensively studied by Jimbo \cite{Jim86}, Woronowicz \cite
{Wor87} and Pasquier\cite{Pas88}. By constructing a q-analogue of the quantum
harmonic oscillator, Biedenharn \cite{Bie89} and Macfarlane \cite{Mac89}
have generalized to $SU_q$(2), the Schwinger approach to the quantum theory
of angular momentum. The q-deformed algebras have found various applications
to physical situations in nuclear and molecular physics \cite{Ray90}\cite
{Bonat91}\cite{Syu90}. Raychev et al.\cite{Ray90} suggest that quantum
algebra is appropriate for the description of stretching effects in
rotational nuclei and have found good fits of rotational spectra of
even-even rare earths and actinides by using a hamiltonian proportional to
the second order Casimir operator of the quantum algebra SU$_q$(2). The SO$%
_q(4)$ quantum algebra has also been used for the description of q-analogue
of the hydrogen atom\cite{Qin93}. In our earlier work \cite{She92} we
constructed a q-deformed analogue of zero coupled nucleon pair states and
found these to be more strongly bound than the pairs with zero deformation,
when a real valued q-parameter is used. An interesting extension to the
multi-shell case, with deformed zero coupled pairs distributed over several
single particle orbits showed that the deformation essentially simulates the
effective residual interaction. Bonatsos \cite{Bon92} has also shown that
the commutation relations of operators for zero coupled correlated fermion
pairs in a single-j orbit can be satisfied up to first order correction by
suitably defined q-bosons, onto which the fermion pair operators are mapped.
Presently, we construct the excitation operators for pairing vibrations from
deformed zero coupled nucleon pairs as well as deformed zero coupled
quasi-particle pairs and analyze the critical point behaviour of nuclei as
the deformation parameter takes real and imaginary values.

Pairing vibrations \cite{Boh64,Hog61,Bes66} are the collective vibrations of
zero coupled fermion pairs around the Fermi surface. The natural framework
for studying the pairing vibrations is RPA for nonsuperconducting nuclei and
Quasi-particle-RPA for superconducting nuclei. In well known works on
pairing vibrations the main concern has not been the reproduction of
experimental data but the physical content of the model. As such several
calculations are available for test nuclei in addition to those for real
nuclei. It is seen that as the interaction between the pairs becomes
stronger a large number of zero coupled pairs are able to cross the Fermi
surface resulting in a phase transition of the nucleus. At this point the
RPA approximation breaks down. The critical point behavior of a nuclear
system is determined to a large extent by how the interaction between the
pairs is taken into account. Our object is to analyze the characteristics of
pairing vibrations induced by boson operators with $J^\pi =0^{+}$
constructed from deformed zero coupled fermion pairs and deformed zero
coupled quasi-particle pairs in order to have a better insight into the
physical content of the deformation parameter.

We have organised the paper as follows. In section 2, we construct bosons
from deformed zero coupled fermion pairs, set up the RPA equations and
obtain the dispersion relation the solution of which gives boson excitation
energies for non-superconducting nuclei. The energies of $0^+$ states of Pb
isotopes are discussed in section 3 and the calculated ratio of two nucleon
transfer cross-sections for populating the $0^+_2$ and $0^+_1$ states of $%
^{208}$Pb is compared with the experimental data to understand the extent
to which anharmonicities are simulated by the deformation. Section 4 deals
with the construction of deformed quasi-boson operators and obtaining
deformed QP-RPA(quasi-particle Radom Phase approximation) equations. The
formalism of sections 2 and 4 is applied in section 5 to the test case of $%
N=20$ particles in two shells and the results for the energy of the lowest $%
0^+$ state from deformed boson approximation and deformed quasi-boson
approximation compared with the exact results. Conclusions are presented in
section 6. The definitions of normalized deformed fermion pair creation
operator and quasi-particle pair creation operators are given in Appendices
A and B respectively.

\section{Deformed pair-RPA equations}

\hspace{.25in} In Ref. \cite{She92} it has been shown that the creation
operator ${Z_0}$ and the annihilation operator ${{{\overline{Z}}_0}}$ for a
deformed zero coupled nucleon pair in a shell-model orbit j may be expressed
in terms of the generators of quantum group S${U_q}$(2). We can
write(appendix (A)),
\begin{equation}
\label{1}Z_0\,=\,{\frac{{{S_{+}}(q)}}{{\sqrt{\{{\Omega }\}_q}}}}\,\quad
;\quad {\overline{Z}}_0=\,{\frac{{{S_{-}}(q)}}{{\sqrt{\{{\Omega }\}_q}}}}
\end{equation}
\begin{equation}
\label{2}{S_0}={\frac{{(n_{op}-{\Omega })}}{{2}}}
\end{equation}
where $(2j+1)=N=2\,{\Omega }$ and
\begin{equation}
\label{3}{\{}x{\}}_q={\frac{(q^x-q^{-x})}{{(q-q^{-1})}}}
\end{equation}
In the following discussion $q=e^\tau $. The vacuum state for q-deformed
pairs is defined through
\begin{equation}
\label{4}{{\overline{Z}}_0}{|}0{\rangle }=0
\end{equation}

For nuclei with no superconducting solution, we construct the boson creation
operator that links the ground state of the nucleus ${\vert}A,0{\rangle}$ to
the $J^{\pi}=0^+$ eigenstate ${\nu}$ of the (A+2) system that is
\begin{equation}
\label{5}{R_{+}^{\nu}}\,={\sum_{m}}\, {X_{m}^{\nu}}\left({\frac{S_{m+}(q)}{
\sqrt{\{ {{\Omega}_m} \}_q}}} \right) -{\sum_{i}}\, {Y_{i}^{\nu}}\left({\
\frac{{S_{i+}}(q)}{\sqrt{\{ {{\Omega}_i} \}_q}}} \right)
\end{equation}
such that
\begin{equation}
\label{6}{\vert}\,A+2,{\nu}{\rangle}= \,{R_{+}^{\nu}}\,{\vert}\,A,0{\rangle}%
\,\,\,; \,\,\, R^{\nu}\,{\vert}\,A,0{\rangle}=\,0
\end{equation}
We use the indices mn(ij) to refer to single particle levels above(below )
Fermi level. The equation of motion\cite{Rowe70} for the operator $%
R_{+}^{\nu}$ is
\begin{equation}
\label{7}{{\langle}\,A,0{\vert}\,[{\delta}R^{\nu},\,[H,\,{R_{+}^{\nu}}]]\, {%
\vert}\,A,0{\rangle}}= {\hbar}{{\omega}_{\nu}}{{\langle}\,A,0{\vert}\, [{%
\delta}R^{\nu},\,{R_{+}^{\nu}}]]\, {\vert}\,A,0{\rangle}}
\end{equation}
where ${\hbar}{{\omega}_{\nu}}=\, (E_{\nu}(A+2)\,-\,{E_0}(A))$. The
amplitudes for zero coupled pair transfer to orbit $j_m$ and $j_i$ are given
by,
\begin{equation}
\label{8}{X_{m}^{\nu}}=\,{{\langle}\,A+2,{\nu}{\vert}\, {\frac{S_{m+}(q)}{
\sqrt{\{ {{\Omega}_m} \}_q}}} {\vert}\,A,0{\rangle}}\,\,;\,\, {Y_{i}^{\nu}}%
=\,{{\langle}\,A+2,{\nu}{\vert}\, {\frac{S_{i+}(q)}{\sqrt{\{ {{\Omega}_i}
\}_q}}} {\vert}\,A,0{\rangle}}
\end{equation}

For the case of independent particles interacting via a pairing force the
system Hamiltonian is
\begin{equation}
\label{9}H={{\sum }_r}{{\epsilon }_r}{{n^r}_{op}}+{H_P},
\end{equation}
where $H_P$ the pairing interaction between deformed pairs is given by
\begin{equation}
\label{10}{H_P}=-G{\sum }_{r,s}S_{r+}(q)S_{s-}(q)
\end{equation}
G being the pairing interaction strength parameter. This is a very simple
model in which interaction strength between pairs is assumed to be the same
irrespective of the j-value of the orbit that the pairs occupy. For this
simple form of the Hamiltonian, the RPA equations are readily obtained by
using the quocommutation relations Eqs.(\ref{6ap1}) for evaluating the
equation of motion, Eq.(\ref{7}). For the two kinds of possible variations, $%
{\delta }R^{\nu}={\frac{S_{n-}(q)}{\sqrt{\{{{\Omega }_n}\}_q}}}$ and ${%
\delta }R^{\nu}={\frac{S_{j-}(q)}{\sqrt{\{{{\Omega }_j}\}_q}}}$, we obtain
the following set of equations for the nucleus (A+2),
\begin{equation}
\label{11}({\hbar }{\omega }_\nu -2{\epsilon }_n)X_n^\nu =-G{\sqrt{\{{{%
\Omega }_n}\}_q}}{\sum_m}\,{X_m^\nu }{\sqrt{\{{{\Omega }_m}\}_q}}-G{\sqrt{\{{%
{\Omega }_n}\}_q}}{\sum_i}\,{Y_i^\nu }{\sqrt{\{{{\Omega }_i}\}_q}}
\end{equation}
\begin{equation}
\label{12}({\hbar }{\omega }_\nu -2{\epsilon }_j)Y_j^\nu =G{\sqrt{\{{{\Omega
}_j}\}_q}}{\sum_m}\,{X_m^\nu }{\sqrt{\{{{\Omega }_m}\}_q}}+G{\sqrt{\{{{%
\Omega }_j}\}_q}} {\sum_i}\,{Y_i^\nu }{\sqrt{\{{{\Omega }_i}\}_q}}
\end{equation}
with solutions
\begin{equation}
\label{13}X_n^\nu =-\frac{N^\nu {\sqrt{\{{{\Omega }_n}\}_q}}}{({\hbar }{%
\omega }_\nu -2{\epsilon }_n)}\quad \, \quad Y_j^\nu =\frac{N^\nu {\sqrt{\{{{%
\Omega }_j}\}_q}}}{({\hbar }{\omega }_\nu -2{\epsilon }_j)}.
\end{equation}
$N^\nu $ defined as
\begin{equation}
\label{14}N^\nu =G{\sum_m}\,{X_m^\nu }{\sqrt{\{{{\Omega }_m}\}_q}}+G{\sum_i}%
\,{Y_i^\nu }{\sqrt{\{{{\Omega }_i}\}_q}}
\end{equation}
evaluated by using the normalization condition,
\begin{equation}
\label{15}{\sum_n}{|X_n^\nu |}^2-{\sum_j}{|Y_j^\nu |}^2=1
\end{equation}
is given by
\begin{equation}
\label{15a}{N^\nu }=\left[ {\sum_n}{\frac{\{{{\Omega }_n}\}_q}{\left( 2{{%
\epsilon }_n}-{{{\hbar }{\omega _\nu}}}\right)^2 }}-{\sum_j}{\frac{\{{{%
\Omega }_j}\}_q}{\left( 2{{\epsilon }_j}-{\hbar }{\omega }_\nu \right)^2 }}%
\right] ^{-\frac 12}.
\end{equation}
The dispersion relation obtained by summing up the Eqs. (\ref{11}) and (\ref
{12})
\begin{equation}
\label{16}{\frac 1{{G}}}={\sum_n}{\frac{\{{{\Omega }_n}\}_q}{{{\left( 2{{{%
\epsilon }_n}-{\hbar }{\omega }_\nu }\right) }}}}-{\sum_j}{\frac{\{{{\Omega }%
_j}\}_q}{\left( {2{{\epsilon }_j}-{\hbar }{\omega }_\nu }\right) }}
\end{equation}
readily yields a graphical solution. A similar equation is obtained for the
eigenstates of the nucleus $(A-2)$ using a two-hole boson creation operator
of the form,
\begin{equation}
\label{17}{R_{+}^\mu }\,={\sum_m}\,{X_m^\mu }\left( {\frac{S_{m-}(q)}{\sqrt{%
\{{{\Omega }_m}\}_q}}}\right) -{\sum_i}\,{Y_i^\mu }\left( {\frac{{S_{i-}}(q)
}{\sqrt{\{{{\Omega }_i}\}_q}}}\right) \,\quad ;\,\quad [H,{R_{+}^\mu }]={%
\hbar }{\omega }_\mu {R_{+}^\mu }
\end{equation}
with the two-hole phonon states given by$\left| (A-2),\mu \right\rangle
=\,\,R_{+}^\mu {|}\,A,0{\rangle .}$

The excited $0^{+}$states of the nucleus ${{|}\,A,0{\rangle }}$ are the
two-phonon states
\begin{equation}
\label{18}{{|}\,\nu ,\mu {\rangle }}=\,R_{+}^\nu \,R_{+}^\mu {|}\,A,0{%
\rangle }
\end{equation}
with excitation energy
\begin{equation}
\label{19}E(0^{+})={\hbar }{\omega }_\nu +{\hbar }{\omega }_\mu
\end{equation}
The operator for two nucleon transfer is
\begin{equation}
\label{20}F=\,{\sum_m}S_{m+}(q)+{\sum_i}S_{i+}(q)
\end{equation}
The amplitudes for populating the ground state and the excited states of the
nucleus ${{|}\,A,0{\rangle }}$, via two nucleon transfer reactions are
\begin{equation}
\label{20a} {{\langle }A,0{|}}FR^\mu_+ {{|}\,A,0{\rangle }}=\frac{N^\mu }G
\end{equation}
and
\begin{equation}
\label{20b} {{\langle }\,\nu,{\mu}^\prime{|}}FR^\mu_+ {{|}A,0{\rangle }}%
=\delta_{{\mu}{\mu^\prime}}\frac{N^\nu }G
\end{equation}
respectively. Here $\mu$ refers to the lowest energy boson.

\section{$0^{+}$ states of Pb Isotopes}

\hspace{.25in}Pb isotopes are a well known and much studied example of
pairing vibrations in nonsuperconducting nuclei. Presently we concentrate on
the $0^{+}$ states of the nucleus $^{208}Pb.$ The main interest is to
calculate the energy of the double pairing vibration(DPV) state for the
neutron pair vibrations in $^{208}$Pb which is experimentally known to lie
at 4.87 MeV with respect to ground state and has been subject of
investigation in various theoretical works on pairing vibrations. Bes and
Broglia\cite{Bes66} using a simplified model consisting of like particles
interacting via a pairing force with constant matrix element predicted an
excited $0^{+}$ state at 4.9 MeV. The calculated ratio between the matrix
elments populating the first excited state and the ground state via a (t,p)
reaction in this model is 1.3. Broglia and Riedel\cite{Brog67} further
anlysed the $^{206}Pb(t,p)^{208}Pb$ data\cite{EXPT1} and showed that the
linear pairing model produces only a qualitative agreement with the
experimental data. Sorensen also \cite{Sorensen67} investigated the neutron
pairing vibrations in $^{208}Pb$ using a pairing force hamiltonian with
constant matrix elements. His model however uses the idea that the
collective excitations of the system can be understood in terms of the
interaction between a few colective bosons, each of which can be expanded in
terms of two-fermion states. The hamiltonian is expressed in terms of the
collective bosons and diagonalized in appropriate collective boson vector
space to obtain the properties of the $0^{+}$ states. In the case of $0^{+}$
states of $^{208}Pb$ , the effect of including anharmonicities in this way
is that though the excitation energy of $0^{2+}$ state is affected only
slightly, the ratio $\frac{\sigma (0_2^{+})}{\sigma (0_1^{+})}$ becomes
closer to the experimental value of 0.45. Sorensen \cite{Sorensen67} points
out that the inclusion of anharmonic terms causes an increase in the (t,p)
cross section for going to the $0_1^{+}$ state of $^{208}Pb$ and a decrease
in the cross section for going to $0_2^{+}$ state. Figures (1a and 1b) show
a plot of left hand side of Eq.(\ref{16}) as a function of the pairing
vibration state energy $E={\hbar }{\omega }$ for two-neutron particle(hole)
addition to the target nucleus $^{208}$Pb. The model space consists of six
hole levels and seven particle levels, the Fermi level being $2g_{\frac 92}$%
. Experimental single particle energies have been used in the calculation.
In figure(1a), we have $q=e^\tau $ with the real valued parameter $\tau $
taking values $0.1,0.2,0.3$ and $0.4$ respectively. The solid line curve
corresponds to undeformed zero coupled pair calculation. We note that as the
deformation increases, the transition to superconducting phase is seen to
occur at progressively smaller values of interaction strength, $G_c$. This
result is consistent with our earlier$\cite{She92}$ conclusion that the real
deformation simulates attractive residual interaction between the nucleons.
For a purely imaginary deformatio n parameter $\tau $ with values i0.1,
i0.2, i0.3 and i0.4 an opposite effect comes into evidence in fig.(1b) where
$G_c$ is seen to become larger as the deformation is increased. The
imaginary deformation $\tau $ effectively decreases the binding energy of
the pair. The variation of $G_c$ versus $|\tau |$ has been plotted in figure
(2) for real as well as imaginary values of $\tau $ .

As in earlier linear calculations \cite{Bes66}\cite{Sorensen67}the energy of
this state is well reproduced for G=0.087 in figure (1a). We notice however
that in the same figure, energy eigen value of 4.87 MeV is reproduced for
DPV also by using deformed pairs with different values of deformation
parameter and corresponding G-values. To choose the deformation parameter
that correctly simulates the anharmonicities, we examine the the behaviour
of two-neutron transfer amplitude F versus boson energy for populating the
ground state $0_1^{+}$ and the DPV state $0_2^{+}$ of $^{208}$Pb for
different values of deformation parameter in figures (3a, and 3b). For
phonon excitation energies close to the unperturbed energy, the F value is
almost independent of the value of parameter ${\tau }$ for two nucleon
addition to $0_2^{+}$. However in the region close to breakdown F is very
sensitive to deformation. We may observe that in general for a given value
of boson excitation energy a real valued deformation causes the (t,p) cross
section for going to $0_1^{+}$ state to increase and the cross section for
going to $0_2^{+}$ state to decrease. These results are very similar to
those from more complex calculations of Sorensen\cite{Sorensen67} showing
that the inclusion of anharmonicities results in an increase in the
calculated value of $\sigma (0_1^{+})$ and a decrease in calculated $\sigma
(0_2^{+})$. An opposite trend is seen in figure (3b) for imaginary values of
$\tau $. Figure (4) is a plot of the ratio $\frac{\sigma (0_2^{+})}{\sigma
(0_1^{+})}$ versus $|\tau |$ ( $0_2^{+}$ being the calculated DPV with
energy 4.87 MeV). For $\tau =0.405$ the experimental value\cite{EXPT1} of $
\frac{\sigma (0_2^{+})}{\sigma (0_1^{+})}=0.45$ is well reproduced.

\section{Superconducting nuclei}

\hspace{.25in} For superconducting nuclei, the operators for creating and
destroying a zero coupled quasi-particle pair and the commutation relations
satisfied by these(in quasi-boson approximation) are given in Eq.({\ref{4ap2}%
}) of appendix B.

In analogy with the case of nonsuperconducting nuclei we expect that the
effect of residual interaction between quasi-particles may be simulated by
deformation of quasi-particle pairs. As such, we define the generators of $%
SU_q$(2) for quasi-particle pairs satisfying the q-commutation relations by
\begin{equation}
\label{21}[{\cal S}_{i-}(q),{\cal S}_{j+}(q)]={\ \{{{\Omega }_i }-{\cal N}_i
\}_q}{\delta _{i ,j }}
\end{equation}
\begin{equation}
\label{22}{\cal N}^i{\cal S}_{i+}(q)=2{\cal S}_{i+}(q)\quad ;\quad {\cal N}^i%
{\cal S}_{i-}(q)=-2{\cal S}_{i-}(q).
\end{equation}

\noindent The creation operator for normalized quasi-particle pair state for
a given j-shell is defined by

\begin{equation}
\label{23}{{\cal Z}^{\dag }}(q){|}0{\rangle }\,=\,{\frac{{{\cal S}_{+}}(q)}{
\sqrt{\{{\Omega }\}_q}}}{|}0{\rangle } .
\end{equation}

Next we construct the quasi-boson creation operator as
\begin{equation}
\label{24}Q_\nu ^{\dag }=\sum_p\left( X_p^\nu {{\cal Z}_p^{\dag }}%
(q)-Y_p^\nu {{\cal Z}_p}(q)\right)
\end{equation}
The ground state of the nucleus $|A,0\rangle $ and the eigenstate $\nu $ of
the nucleus $|A+2,\nu \rangle $ are related by
\begin{equation}
\label{25}{|}\,A+2,{\nu }{\rangle }=\,{Q_\nu ^{\dag }}\,{|}\,A,0{\rangle }%
\,\,\,;\,\,\,Q_\nu \,{|}\,A,0{\rangle }=\,0
\end{equation}
To derive the equations satisfied by amplitudes $X_p^\nu $ 's and $Y_p^\nu $
's we compute the commutators $[\delta Q_\nu,[{\cal H}_{11}+{{\cal H}_c},{%
Q_\nu ^{\dag }}]]$ for the variations $\delta Q_\nu={{\cal Z}_m}(q)$ and $%
\delta Q_\nu={{\cal Z}_m}^{\dag }(q)$ and take their expectation values with
respect to the ground state $|A,0\rangle $ i.e. evaluate,
\begin{equation}
\label{26}\langle A,0|[\delta Q_\nu,[{\cal H}_c,{Q_\nu ^{\dag }}%
]]|A,0\rangle ={\hbar }{\omega }_\nu \langle A,0|[\delta Q_\nu,{Q_\nu ^{\dag
}}]|A,0\rangle .
\end{equation}
Here the Hamiltonian ${\cal H}_{11}+{{\cal H}_c\ }$is obtained by replacing
the normal quasi-particle pair creation and destruction operators in $H_{11}+%
{\ H}_c$ of appendix B by deformed quasi-particle pair creation and
destruction operators that is
\begin{eqnarray}
{\cal H}_{11} & = & {\sum_i}E_i {\cal N}^i%
\\ {{\cal H}_c} &  = & - G{{\left( %
{\sum_{ij} }u_i^2{u_j^2}{\cal S}_{i+}(q){\cal S}
_{j-}(q)-u_i^2 {v_j^2}{\cal S}_{i+}(q){\cal S}_{j+}(q)\right. }}\nonumber
\\  &   & \left. -v_i^2 {u_j^2}{\cal S}_{i-}(q){\cal S}_{j-}(q)+v_i^2
{v_j^2}{\cal S}_{i-}(q){\cal S}_{j+}(q)\right) .
\end{eqnarray}

\hspace{0.25in} In deformed quasi-boson approximation, on evaluating Eq.(\ref
{26}) the amplitudes $X^\nu $ 's and $Y^\nu $ 's are found to satisfy the
following set of coupled equations,
\begin{equation}
\label{27}({\hbar }{\omega }_\nu -2E_m)X_m^\nu =-G\sqrt{\{{{\Omega _m} }\}_q}%
\sum_p\,\sqrt{\{{{\Omega _p} }\}_q}\left( X_p^\nu
(u_m^2u_p^2+v_m^2v_p^2)-Y_p^\nu (u_m^2v_p^2+v_m^2u_p^2)\right)
\end{equation}
\begin{equation}
\label{28}({\hbar }{\omega }_\nu +2E_m)Y_m^\nu =G\sqrt{\{{{\Omega _m} }\}_q}%
\sum_p\,\sqrt{\{{{\Omega _p} }\}_q}\left( Y_p^\nu
(u_m^2u_p^2+v_m^2v_p^2)-X_p^\nu (u_m^2v_p^2+v_m^2u_p^2)\right) .
\end{equation}
The set of Eqs. (\ref{27},\ref{28}) can be solved by using standard methods
to furnish the roots $E={\hbar }{\omega }_\nu$.

\section{Two Shell Test Model}

\hspace{.25in} For analyzing the behaviour of pairing vibration states in
the deformed boson approximation and deformed quasi-boson approximation, we
examine these states for a test model in which $N=2\Omega $ particles are
distributed over two shell model orbits, each with a degeneracy $2\Omega $.
The Shell model orbits have single particle energies given by $\frac
\epsilon 2$ and $-\frac \epsilon 2\ $respectively. Following Hogaasen \cite
{Hog61}, we plot in fig.(5) the lowest energy root in units of $2\epsilon $
as a function of $\frac{G\Omega }{2\epsilon }$. The plot is for $N$=20 and
shows the results for various values of deformation parameter $\tau $ along
with the results for exact calculation, boson approximation with zero
deformation($\tau =0.0)$ and quasi-boson approximation without deformation.
All boson approximation curves start at $\frac E{2\epsilon }=1.0$ and
terminate at $\frac E{2\epsilon }=0.0$ for a $G$ value characteristic of each
deformation value. The deformed quasi-boson approximation curves in general
follow the trend of the similar curve for $\tau =0.0$. In fig.(5a) the
deformation parameter takes the real values $\tau =0.1,0.15$ and $0.2.$ We
may notice that for a given value of pairing interaction strength $G$ the
deformation causes the energy values to be lowered as compared to the boson
approximation results for $G<G_c$ and quasi-boson approximation results for $%
G>G_c$. It implies that a boson or a quasi-boson pair with real deformation
is more strongly bound than its undeformed counterpart. As the quasi-boson
approximation energies lie higher than the exact energy eigen value for a
given $G$, by using a suitable deformation exact energies can be reproduced
in the deformed quasi-boson approximation. In the case at hand, for $\tau
=0.15$ there is a good agreement between the deformed quasi-boson
approximation calculation and the exact result for values of $G$ not very
close to $G_c.$ For deformed boson calculation the critical value of pairing
strength $G$, for which the energy of the lowest energy state approaches
zero, is seen to go up as the deformation is increased. In other words,
increasing the deformation parameter causes the transition to
superconducting phase to occur at successively lower values of $G$.

In fig.(5b) the deformation parameter is taken to be purely imaginary with
the curves shown for $\tau =i0.05,i0.1$ and $i0.15.$ For a given value of $G$%
, an imaginary deformation parameter $\tau $ raises the excitation energy in
the boson as well as the quasi-boson approximation. It may be interpreted as
an anti-pairing effect. As the boson approximation energies lie lower than
the exact energy eigenvalues for the same $G$ value, for a suitable choice
of $\tau $ we can obtain a deformed boson approximation curve having a good
overlap with the exact results for a reasonably large range of $G$ values in
the region away from the phase transition region. Deformation in this case
simulates the correlations not accounted for in the boson approximation. In
a realistic case, an imaginary deformation may be used as a measure of
correlations caused by a residual repulsive interaction not being accounted
for in the boson or quasi-boson approximation.

We have plotted in Fig.(5c) the deformed boson approximation energies for $%
\tau =i0.104$ and quasi-boson energies for $\tau =0.15$ the deformation
values for which a good agreement with the exact calculation results is
obtained in a wide region away from the phase transition region.

\section{Conclusions}

\hspace{.25in} We have constructed q-analogues of boson approximation and
quasi-boson approximation to understand the physical meaning of real and
imaginary deformation in the context of interaction between zero coupled
pairs. The formalism for non-superconducting nuclei has been applied to the
classic example of the Double Pairing Vibrational state $0_2^{+}$ in the
nucleus $^{208}$Pb. For a real deformation value of $\tau =0.405$, the
calculation reproduces the observed excitation energy of the $0_2^{+}$state,
$E=4.87$ MeV, as well as the ratio of cross sections, $R=\frac{\sigma
(0_2^{+})}{\sigma (0_1^{+})}=0.45$, for populating the $0_2^{+}$ and $0_1^{+}
$ states via (t,p) reaction. A linear calculation \cite{Bes66}without
deformation produces a large value of R=1.3. For obtaining a better
agreement with experiment a more complex calculation that takes into account
anharmonic effects as of Ref\cite{Sorensen67} is necessary. However
presently a good agreement with the experiment is obtained in a much simpler
model by using deformed pairs. We may infer that a real deformation of zero
coupled pairs takes into account the anharmonicities not being accounted for
in boson approximation in the simple model at hand with the same pairing
strength for all orbits . We also note that as the deformation parameter is
increased the transition to superconducting phase occurs for successively
smaller values of pairing interaction strength parameter for $\tau $ real
and for sccessively larger values of $G$ for imaginary values of $\tau $.
This behaviour is consistent with our earlier conclusion\cite{She92} that
the deformed fermion pairs are more strongly bound for real valued $\tau $
and the deformation in this case simulates the residual attractive
interaction. Similar results have been obtained by Bonatsos et. al \cite
{Bon93} in the framework of Moszkowski model where increasing q-deformation
has been shown to facilitate the phase transition from the vibrational to
the rotational behavior. For $q=e^{i\tau }$, on the other hand, deforming
the pair amounts to weakening the pair. The role of complex valued q is akin
to a pair breaking residual repulsive force.

In both the cases the collectivity of the state is seen to be a function of
deformation parameter as is evident from the variation of calculated two
nucleon transfer cross sections in figures(3a,3b). We may point out that
deforming the zero coupled pairs is not the same as changing the value of
the pairing strength parameter G. For phonon excitation energies close to
the unperturbed energies the collectivity is almost independent of the value
of parameter ${\tau }$. However in the region close to phase transition it
is very sensitive to deformation. The deformed pair Hamiltonian apparently
accounts for many-body correlations, the strength of higher order force
terms being determined by the deformation parameter .

For the test model of 20 particles in two shells, the results of q-deformed
boson and quasi-boson approxmations have been compared with exact results.
The deformed boson approximation results for $\tau =i0.104$ and deformed
quasi-boson energies for $\tau =0.15$ overlap the exact calculation results
in a wide region away from the phase transition region. Apparently the
deformation effectively takes into account the correlations not being
accounted for in approximate treatments. As such, in a realistic calculation
the deformation parameter may be taken as a quantitative measure of
correlations left over in an approximate treatment as compared to the exact
treatment. The test model results that the deformed boson pairs and deformed
quasi-boson pairs are more(less) strongly bound as compared to their
undeformed counterparts, for real(imaginary) $\tau$, confirms  our earlier
conclusions.

We may conclude that, in a realistic case, a real(imaginary) deformation may
be used as a measure of anharmonicities caused by a residual
attractive(repulsive) interaction not being accounted for or a measure of
correlations left over in an approximate treatment.

\section{Acknowledgements}

\hspace{.25in}This work has been partially supported by CNPq, Brazil .

\appendix

\section{Normalized Deformed pair creation operator}

\hspace{.25in} We can write a zero-coupled pair for two nucleons \cite{Fre66}
in a shell model orbit j as
\begin{equation}
{Z_{0}}=-{\frac{1}{{\sqrt{2}}}}\,({A^{j}}\times{A^{j}})^0 \nonumber
\end{equation}
with
\begin{equation}
{\overline{Z_{0}}}={\frac{1}{{\sqrt{2}}}}\,({B^{j}}\times{B^{j}})^0
\nonumber
\end{equation}
where
\begin{equation}
{A_{jm}}= {a^{\dag}}_{jm}\quad ;\quad {B_{jm}}=(-1)^{m+j}{a_{j,-m}}
\nonumber
\end{equation}

The fermion creation and destruction operators ${{a^{\dag}}_{jm}}$ and $%
a_{jm}$ satisfy the usual commutation relations
\begin{equation}
[{{a^{\dag}}_{jm}},{a_{jm}}]=1 \nonumber
\end{equation}
And the number operator is defined as
\begin{equation}
{n_{op}}={\sum_{m}}\, {{a^{\dag}}_{jm}}{a_{jm}} \nonumber
\end{equation}
We can easily verify that
\begin{equation}
[{Z_0},{\overline{Z}}_0]={\frac{{n_{op}}}{{\Omega}}}-1 \nonumber
\end{equation}
\begin{equation}
\label{1ap1}[{n_{op}},{Z_{0}}]=2\,{Z_0}\quad \quad [{n_{op}},{\overline{Z}}%
_{0}]=-2\,{\overline{Z}}_{0}.
\end{equation}
Here $(2j+1)=N=2\,{\Omega}$. We may rewrite our pair operators in terms of
the well known quasi-spin operators by identifying,
\begin{equation}
{{S_{+}}}={\sqrt{\Omega}}\,Z_{0}\quad \quad {{S_{-}}}={\sqrt{\Omega}}\,{\
\overline{Z}} _{0} \nonumber
\end{equation}
\begin{equation}
\label{2ap1}{{S_{0}}}={\frac{{(n_{op}-{\Omega})}}{{2}}}.
\end{equation}
${{S_{+}}}$, ${{S_{-}}}$ and ${S_{0}}$ are the generators of Lie algebra of
SU(2) and satisfy the same commutation relations as the angular momentum
operators.
\begin{equation}
\label{3ap1}[{{S_{+}}},{{S_{-}}}]=2{{S_{0}}} \quad \quad [{{S_{0}}},{{S_{
\underline{+}}}}]= {\underline{+}}{{S_{\underline{+}}}}
\end{equation}
Total quasi-spin operator is given by
\begin{equation}
\label{4ap1}{S^2}\,=\,{{S_{+}}}{{S_{-}}}\,+ {{S_{0}}}({{S_{0}}}-1)
\end{equation}

An equivalent description of the state ${\vert}n, v{\rangle}$ can be given
in terms of the total quasi-spin quantum number $s$ (related to the
seniority quantum number $v$ through $s={\frac{({\Omega}-v)}{2}}$) and the
eigenvalue of operator ${S_0}$. The states ${\vert}s, s_0{\rangle}$ satisfy
the following relations,
\begin{equation}
\label{5ap1}{S}^2\,{\vert}s, s_0{\rangle}\,=\,s(s+1)\,{\vert}s, s_0{\rangle}%
\quad; \quad{S_0}{\vert}s, s_0{\rangle}\,=\,{s_0}\,{\vert}s, s_0{\rangle}
\end{equation}

We may now define q-deformed pairs in terms of the generators of ${SU_q}(2)$
satisfying the commutation relations,
\begin{equation}
\label{6ap1}[{{S_{+}}}(q),{{S_{-}}}(q)]=\{2{{S_{0}}}(q)\}_q \quad \quad [{%
S_{0}}(q),{S_{\underline{+}}}(q)]={\underline{+}}S_{\underline{+}}(q)
\end{equation}
where
\begin{equation}
\label{7ap1}{\{}x{\}}_{q} ={\frac{(q^x-q^{-x})}{{(q-q^{-1})}}}
\end{equation}

Normalized two nucleon state in terms of the deformed pair creation operator
is
\begin{equation}
\label{8ap1}{Z_0}{\vert}0{\rangle}\,=\, N\, {{S_+}(q)}= \, {\frac {{S_+}(q)}
{\sqrt{\{ {\Omega} \}_q}}}{\vert}0{\rangle}
\end{equation}
As such the operator for the creation of a normalized deformed pair state is
${\frac{{S_+}(q)}{\sqrt{\{ {\Omega} \}_q}}}$.

\section{Quasi-particle pair creation and annihilation operators}

\hspace{.25in} The creation and destruction operators for quasi-particles
are defined in terms of fermion creation and destruction operators by
\begin{equation}
\label{1ap2}\alpha _{jm}^{\dag }=u_ja_{jm}^{\dag }-(-1)^{j-m}v_ja_{j-m}\quad
;\quad \alpha _{jm}=u_ja_{j-m}-v_ja_{jm}^{\dag }\quad ;\quad u_j^2+v_j^2=1
\end{equation}
Quasi-spin operators for quasi-particles in an orbit j,

\begin{equation}
\label{2ap2}{{\cal S}_+}=\, \sum_{m>0} (-1)^{j-m}
\alpha_{jm}^{\dag}\alpha_{j -m}^{\dag}\quad;\quad {{\cal S}_-}=\, \sum_{m>0}
(-1)^{j-m} \alpha_{j -m}\alpha_{jm}
\end{equation}
and the quasi-particle number operator defined as
\begin{equation}
\label{3ap2}{\cal N}= \sum_m \alpha_{jm}^{\dag}\alpha_{jm}
\end{equation}
satisfy the following commutation relations
\begin{equation}
\label{4ap2}\left[ {{\cal S}_{i-}},{{\cal S}_{j+}}\right]= \left({\Omega_i}-
{{\cal N}^i}\right)\delta_{ij}\quad;\quad \left[ {\cal N}^i,{{\cal S}_{j {
\underline{+}}}}\right]= {\underline{+}}2 {{\cal S}_{i{\underline{+}}}}%
\delta_{ij}
\end{equation}
Normalized zero coupled quasi-particle pair creation operator for a given
orbit j is given by,
\begin{equation}
\label{5ap2}{{\ Z}^{\dag}}{\vert}0{\rangle}\,= \, {\frac {{{\ S}_+}} {\sqrt{%
\{ {\Omega} \}}}}{\vert}0{\rangle}
\end{equation}

The pairing Hamiltonian in terms of these operators is written as
\begin{equation}
H=H_{00}+H_{11}+H_{20}+H_{02}+H_c+H_{res}\nonumber
\end{equation}
Where
\begin{eqnarray}
	H_{00} & = & \sum_i ( \epsilon_i- \lambda) 2 v_i^2 \Omega_i
+G( \sum_i \Omega_i u_i v_i )^2 \nonumber \\
	H_{11} & = &  \sum_i [( \epsilon_i- \lambda)
   (u_i^2-v_i^2)+2Gu_i v_i
( \sum_j \Omega_j u_j v_j )^2]{\cal N}_i  \nonumber \\
H_{20} +H_{02} & = & \sum_i [ ( \epsilon_i- \lambda) 2 u_i v_i -
G \sum_j \Omega_j u_j v_j (u_i^2-v_i^2)]({\cal S}_{i+} +{\cal S}_{i-})
\nonumber \\
	   H_c & = & -G \sum_{ij}( u_i^2 u_j^2 {\cal S}_{i+}
		     {\cal S}_{j-}
		     -u_i^2 v_j^2 {\cal S}_{i+} {\cal S}_{j+} \nonumber \\
	       &   & -v_i^2 u_j^2 {\cal S}_{i-} {\cal S}_{j-}
		     +v_i^2 v_j^2 {\cal S}_{i-} {\cal S}_{j+})
		\nonumber\\
       H_{res} & = & G \sum_{ij}u_i v_i [ {\cal N}_i
		    (u_j^2 {\cal S}_{j-}
- v_j^2 {\cal S}_{j+}) + ( u_j^2 {\cal S}_{j+} - v_j^2 {\cal S}_{j-} )
{\cal N}_i ] \nonumber \\
	       &   & -G \sum_{ij} u_i v_i u_j v_j {\cal N}_i {\cal N}_j.
\label{6ap2}
\end{eqnarray}

Imposing the condition that $H_{02}$ and $H_{20}$ cancel each other and the
vacuum state for quasi-particles represent the ground state of the nucleus
with n nucleons, we obtain the well known expressions for occupancies and
the gap parameter.
\begin{eqnarray}
u_{i}^2 & = &\frac{1}{2} \left( 1+
\frac{{{\epsilon}_i}-{\lambda}}{E_i}
\right) \quad ; \quad  v_i^2
={1\over2}\left(1-\frac{{\epsilon}_i-{\lambda}}{E_i}\right)
  \nonumber \\
\sum_i \frac{\Omega_i}{E_i} & = & \frac{2}{G}\quad;\quad
{\sum_i} {\Omega_i}\left(1- \frac{\epsilon_i-\lambda}{E_i}
\right)=n \nonumber \\
E_i & = & \sqrt{\Delta^2+({\epsilon_i-\lambda})^2}\nonumber\\
\Delta & = & G {\sum_i} {\Omega_i}{u_i} {v_i}
\label{7ap2}
\end{eqnarray}

\newpage
\begin{titlepage}
\begin{center}
\LARGE{Figure Captions}
\end{center}

Fig. (1a) - {A plot of left hand side of equation (17) as a function
of $E={\hbar}{\omega}$ for $q=e^{\tau}$ with $\tau$ real. The
vertical straight lines are
$2{\epsilon}_{3p_{1\over2}}= -2.7$ MeV,
$2{\epsilon}_{3g_{9\over2}}=4.15$ MeV, ${\hbar}{\omega}_\nu= -2.2$ MeVand
${\hbar}{\omega}_\mu= 2.65$ MeV.}

\vspace{0.8 true cm}
Fig. (1b) - As in (1a) for $\tau$ imaginary.

\vspace{0.8 true cm}
Fig. (2) - The breakdown point value of pairing interaction strength
$G_C$ as a function of ${|\tau |}$ for $\tau$ real as well as imaginary.

\vspace{0.8 true cm}
Fig. (3a) - Two nucleon transfer amplitude for populating the states
$0^+_1$ and $0^+_2$ of $ ^{208}$Pb as a function of E
for $\tau$ real.

\vspace{0.8 true cm}
Fig. (3b) - As in (3a) for $\tau$ imaginary.

\vspace{0.8 true cm}
Fig. (4) - The ratio of cross-sections for populating the states
$0^+_2$ and $0^+_1$ via two-neutron transfer in $ ^{208}$Pb, $\frac{\sigma
(0^+_2)}{\sigma (0^+_1)}$ versus $|\tau|$ ( $0^+_2$ is the calculated DPV state
with energy 4.87 MeV) for $\tau$ real as well as imaginary.

\vspace{0.8 true cm}
Fig. (5a) -The lowest energy root in units of $2\epsilon $ plotted as a
function of $\frac{G\Omega }{2\epsilon }$. The plot for $N$=20 shows
the results for deformed boson and quasi-boson
approximations for $\tau$ real along with
the  exact calculation, boson approximation and
 quasi-boson approximation without deformation.

\vspace{0.8 true cm}
Fig. (5b) - As in (5a) for $\tau$ imaginary.

\vspace{0.8 true cm}
Fig. (5c) - As in (5a) for $\tau=i0.104$ in deformed boson
approximation and for $\tau=0.15$ in deformed quasi-boson approximation.
\end{titlepage}

\newpage
\begin{titlepage}
Keyword Abstract: Quantum group ${SU_{q}}(2)$, Pairing-interaction, q-deformed
RPA and
QRPA,  ${0^+}$  States of $ ^{208}$Pb, Two
neutron transfer cross-section, Test model of 20 particles in two
shells
\end{titlepage}

\end{document}